\documentclass[a4paper,fleqn,usenatbib]{mnras}

\usepackage{newtxtext,newtxmath}
\usepackage[T1]{fontenc}
\usepackage{ae,aecompl}

\usepackage{graphicx}	
\usepackage{amsmath}	
\usepackage{amssymb}	
\usepackage{color}
\newcommand{\gc}{$\gamma$\,Cas}
\newcommand{\gC}{$\gamma$\,CAS}
\newcommand{\xr}{X-ray}
\newcommand{\XR}{X-RAY}

\def\gtrsim{\mathrel{\hbox{\rlap{\hbox{\lower4pt\hbox{$\sim$}}}\hbox{$>$}}}}
\def\ltsim{\mathrel{\hbox{\rlap{\hbox{\lower4pt\hbox{$\sim$}}}\hbox{$<$}}}}
\definecolor{gray}{rgb}{0.5,0.5,0.5}

\title{Is there a propeller neutron star in $\gamma$\,Cas?}

\author[M. A. Smith, R. Lopes de Oliveira, \& C. Motch]{
M. A. Smith,$^{1}$\thanks{E-mail: myronmeister@gmail.com }
R. Lopes de Oliveira,$^{2}$
C. Motch$^{3}$   \\
$^{1}$National Optical Astronomy Observatory, 950 N. Cherry Ave., Tucson, AZ, 
USA\\
$^{2}$Universidade Federal de Sergipe, Departamento de F\'isica, Av. Marechal 
Rondon, S/N, 49000-000 S\~ao Crist\'ov\~ao, SE, Brazil; \\
Observat\'orio Nacional, Rua Gal. Jos\'e Cristino 77, 20921-400, Rio de
Janeiro, RJ, Brazil \\
$^{3}$Universit\'e de Strasbourg, CNRS, Observatoire Astronomique, UMR 7550, 
F-67000, Strasbourg, France
}

\date{Accepted XXX. Received YYY; in original form ZZZ}

\pubyear{2015}

\begin{document}
\label{firstpage}
\pagerange{\pageref{firstpage}--\pageref{lastpage}}
\maketitle

\begin{abstract}
$\gamma$\,Cas is the prototype of a small population of B0-B1.5 III-V classical
Be (cBe) stars that emit anomalous and hard X-rays with a unique array of 
properties. \gc\ is known to host, like other cBe stars, a decretion disk
and also a low mass companion.
Recently Postnov et al. have posited that this companion is a magnetized 
rapidly spinning neutron star that deflects direct gravitational accretion 
from a stellar/disk wind via the ``propeller mechanism." These authors 
state that the key X-ray observations are ``remarkably well produced" in this
scenario.  We reexamine this mechanism in detail and conclude that there 
are a number of fatal objections in its application to the \gc\ case. 
Among other considerations these issues include the prediction under the 
propeller scenario of a 
much smaller population of \gc\ stars than is observed and the lack of
allowance for observed correlations of X-ray and UV and/or optical properties 
over a variety of timescales.
\end{abstract}



\section{Introduction}

 Over the last twenty years we authors 
have conducted a number of observational campaigns to monitor the properties 
in the X-ray, optical, and ultraviolet domains of the 
hard X-ray emitter \gc\ (B0.5\,IV-Ve). All this work has led
us to a picture in which these anomalous \xr\ emissions are 
produced by the interaction of magnetic fields of this star and its Be 
``decretion" disk. This holds as well among ``analog" members of this new 
X-ray Be subgroup. 

In this picture X-rays are generated from high energy particle beams 
directed to the surface of the Be star and depositing their energy as hard 
X-ray-visible quasi-flares. These beams are created by acceleration 
of particles situated within the ambient field lines following their
entanglement of two magnetic field structures. 
The first structure consists of local chaotic field lines emanating 
from and corotating with the star. 
The second is a toroidal field embedded in the Be star's Keplerian disk and
perhaps amplified by the \citet[][]{Balbus&Hawley1991} magnetorotational 
instability \citep[][``RSH02'']{Robinson et al.2002}. 
\citet[][``RS00'']{Robinson&Smith2000} reported numerical simulations 
to show what attributes the high energy electron beam might have to 
generate the observed \xr\ flux. Of course there is no way to directly 
observe this complicated interplay. 
Rather it can be inferred only as a result of employing a variety of observational techniques.

In a recent paper \citet[][``POT17'']{Postnov et al.2017} have
proposed instead that the hard \xr\ emission from the \gc\ system is 
caused by the operation of 
a ``propeller" associated with a rigidly rotating magnetosphere around 
a {\it putative neutron star} (NS) secondary.
an idea first raised in the context of \gc~ long ago by \citet[][]{Corbet1996}.
However, this narrative has intrinsic weaknesses and overlooks key
observational properties of the Be star and disk. We believe also the 
mechanism they propose is inherently untenable for these stars, and we 
address these issues herein. In doing so we occasionally draw 
on additional information known from other analogs of this \gc\ 
subgroup. Much of this information is taken from a general review paper 
\citep[][``SLM16'']{Smith et al.2016} and references cited therein.

\section{\XR\ and related properties of \gC}
\label{reltd}

  \gc\ is the prototype of a subgroup of 10-12 Galactic \xr\ classical 
Be stars. Members of this subgroup  are confined to the region B0-B1.5, 
luminosity class III, IV in the HR Diagram (SLM16). RV studies of \gc\ 
reveal it to be a single-line binary in a wide, circular orbit 
(P = 203.59 days; $e$ $\le$ 0.03  \citep[][``SLM12a'']{Smith et al.2012a}. 
The secondary's mass is about 0.8\,$\pm{0.4}$ 
M$_{\odot}$. Otherwise, we stress that its evolutionary status is unknown. For
example, the secondary could be a late-type main sequence star or for
that matter a passive degenerate  star.
Long Baseline Optical Interferometry from a number of studies 
has fixed the obliquity of the Be-disk system to our line of sight as 
$i$\,$\approx$ 42$^{o}$ (SLM12a).
The Be component of \gc\ has a rotational period of $\approx$1.22 days
\citep[][``HS12'']{Henry&Smith2012}. 
This conclusion comes from a 
robust periodic feature in its optical light curve and is consistent with the 
extreme rotational broadening of its spectral lines. Given this period, its 
estimated radius and sin\,i obliquity, \gc\ appears to rotate nearly at
the critical velocity. This rotation could produce localized magnetic 
dynamo instabilities due to an enhanced equatorial convection zone 
\citep[][``MLS15'']{Motch et al.2015}.
This may speak to the cause of the star's role 
in generating hard X-rays and can address the possible criticism that most 
other early Be stars with extensive disks do not emit this radiation. 
Likewise, it is sometimes speculated that \gc\ is in an 
intermediate state of binary evolution because as many as three members
of the subgroup could be blue stragglers in clusters (SLM16). 

  \gc\ stars are recognizable by their unique array of
\xr\  emission characteristics among high mass X-ray stars. 
The emission is "hard," and their moderate
$L_{\mathrm{x}}$ luminosity, 3-10$\times$10$^{32}$ ergs\,s$^{-1}$,
(as measured in the 0.1-10 keV \xr\ band) is intermediate between so-called 
classical Be stars and BeXR binary systems. High resolution spectra of  
$\gamma$\,Cas, which include the continuum and resolve the Lyman\,$\alpha$ 
emission lines of Fe XXV (at 6.7\,keV) and Fe\,XXVI (at 6.9 keV), are 
fit well with an optically 
thin thermal model described by a ``primary" plasma temperature 
``$kT_{\rm hot}$" $\approx$14\,keV and (usually) an attenuation of soft 
\xr\ flux by photoelectric absorption of ``cold" matter. 
Importantly, for \gc\ a second (often higher) column density of cold matter is
also present in the foreground of the hot primary \xr\ plasma. We should note 
clearly that the flux of the primary component dominates at all \xr\ wavelengths
(save the wavelengths of soft and moderate energy \xr\ lines).  It follows 
that the attenuation of soft X-ray signals the presence of this cold matter 
along the line of sight to the hot plasma behind it. It remains to be 
added that fluorescence lines of Fe and Si are present in the \gc\ \xr\ 
spectrum. These features, which change with time, indicate the presence of 
cold matter close to the hot \xr\ sources. 

The primary plasma can take on various $kT_{\rm hot}$ values for different
members of the \gc\ subgroup. For at least half of those \gc\ stars 
observed more than once the primary plasma's temperature has been found
to change.  In addition, for the best observed stars the presence of 
emission lines of H- and He-like lines of lower ionization metallic 
species  at lower energies (0.3-0.6 keV band) discloses that 2-3 secondary 
cooler plasmas are present. These plasmas seem to be monothermal because
a model described by a Differential Emission Measure, that is, by a plasma 
having by a continuous distribution of temperatures, does not give a good fit
\citep[][]{Lopes de Oliveira et al.2010}, (SLM12a).

Although the abundances derived from 
spectral analysis are generally consistent with solar values, for a 
few elements their values are nonsolar and these abundances vary over time. 
These include over or underabundances of atomic neon and nitrogen. These anomalies are inconsistent with abundances produced by core or shell burning in late stages of 
stellar evolution, e.g., of a putative pre-supernova.
Curiously, a persistent low iron abundance has been found from the Fe K-shell 
lines of Fe\,XXV and Fe\,XXVI determined by many
authors using as many X-ray telescopes, even though the abundance derived 
from L-shell ion lines is solar-like.

 The X-ray light curve of \gc\ exhibits an array of variations 
characterized by certain timescales. These include (1) ubiquitous 
``shots" or quasi-flares lasting a (few seconds to a minute), 
(2)  erratic ``undulations" over 10s of minutes to hours, 
(3) long-cycles ($\sim$70 days), and 
(4) seemingly chaotic, even longer term variations (SLM16, MLS15). 
In addition to the ``flares," variations of an underlying ``basal" X-ray flux 
component can be present. This component constitutes $\approx$$\frac{2}{3}$ 
of the total flux, exhibits the same $kT_{\rm hot}$ as the flares, and 
varies on a timescale of a few hours (bullet 2).  
Unlike true magnetic flares in cool stars, the tails of the flare profiles 
exhibit no tapering or extensions \citep[][``SRC98'']{Smith et al.1998a}. 
This fact suggests a rapid decay timescale. 
In general, the flares are visible in both soft and hard \xr\ bandpasses,
but exceptions occur in which they are present in one energy band and
are weak or absent in the other according to \citet[][``SLM12b'']{Smith 
et al.2012b}. This study also found that the flare properties of \gc\ are
essentially the same for another \gc\ analog, HD\,110432 (B0.5\,IIIe).

Power spectra of the short timescales (1) and (2) variations describe epoch-dependent, mild deviations from a $1/f^{n}$ relation (where $n$ is always near unity). However, the details of these relations depend upon an epoch-dependent 
distribution of flare strengths.

The quasi-sinusoidal $\sim$70-day cycle was first observed in optical 
photometry. This cycle is almost ubiquitous and in the \xr\ region 
attain an amplitude of a factor of 2-3.   These \xr\ variations
generally correlate well with optical variations (RS00, HS12, MLS15). 
The Johnson $V$ band amplitudes of the optical cycles are 
greater than the $B$ ones. This means the only other possible contributor
to red-optical flux in the Be complex, the Be decretion disk, must 
play an important role in the generation of at least the variable component 
of the hard \xr\ flux. This
is the primary reason for including the disk in the magnetic interaction
picture.

On a timescale of several months, MLS15 found that optical flux changes 
caused by density variations in the inner part of the decretion disk are 
strongly correlated with X-ray fluxes with a similar flux ratio as seen for 
the 70\,d cycles. They found that optical and X-ray fluxes vary together 
with a time delay of less than one month. These authors have also shown 
that such a small time lag is inconsistent with the expected transit time 
from inner disk regions to the orbit of the companion.

For completion, we note that UV light curves of \gc\ exhibit small dips 
lasting a few hours, and optical helium line profiles exhibit blue-to-red 
rapidly moving migrating subfeatures \citep[][]{Smith et al.1998b}.
Neither of these signatures is correlated with the star's \xr\ flux, 
though both indicate the presence of plasma forced into corotation just 
over the star's surface. 
  
  Apart from the hard flux variations, the soft relative 
to hard energy flux can change on timescales of ten days or longer (SLM12a), 
and at times rapidly over several minutes \citep[][]{Hamaguchi et al.2016}. 
This is the ``attenuation effect" noted above. The occurrence of this effect 
is important because it conveys important geometrical details of the hard
\xr\ sources. In this case SLM12a noted that the presence of intervening
cold matter indicates no relation with binary phase of \gc\ during an 
outburst event in 2010. 
Significantly, the cold matter absorption column increased
simultaneously with the beginning of the 2010 Be outburst, which is to say
the ejection of matter from the Be star. The occurrence of this column of
the cold matter places the hot, high plasma density, X-ray sources between 
observer and the Be star.  We develop this point below.

A particularly salient discovery concerning the hard X-ray flux is 
the tight relation between its variations and ultraviolet line 
strengths and continuum, and optical light curves (SLM16 and references cited
therein). For example, from a 21-hour simultaneous monitoring in 1996 of \gc\ 
by the {\it Rossi X-ray Telescope Explorer (RXTE)} and the {\it Goddard High 
Resolution Spectrograph (GHRS)} attached to the {\it Hubble Space Telescope,} 
\citet[][``SR03'']{Smith&Robinson2003} reported 
that increases in \xr\ flux were
associated with strengthenings of a UV Fe\,V line and weakenings of 
UV Si\,III lines on timescales of several minutes. 
This behavior is consistent only with plasma near a bright
and optically thick medium, namely when the UV line strengths increase or
decrease in response to
a closeby \xr\ source. 
These relationships constitute a particularly critical argument in locating 
the X-ray plasma near the Be star. 

  We underscore that observations reported in many studies of 
early-B/degenerate star binaries at large disclose that ultraviolet light from
the secondary component is not visible against the primary's contribution. 
Even for systems with hot degenerate secondaries essentially all the 
UV flux comes from the B star.  Also, while initial hydrodynamical simulations 
suggest that the disks are likely to be truncated 
by 3:1 orbital tidal resonances \citep{Okazaki&Negueruela2001}, the actual
mass transfer rate deposited to the secondary (whether degenerate or 
otherwise) is unknown.  

\section{Properties of X Per}

In the POT17 scenario \gc\ can evolve to become a X\,Per-like object. 
Therefore we summarize the relevant attributes of this well-known system. 
The Be binary \xr\ pulsar
X\,Per/4U352+309 consists of an O9\,III primary star 
and X-ray pulsing (P$_{\rm puls}$ $\sim$ 837 s) NS. As summarized by 
\citet[][``LTC12'']{Lutovinov et al.2012}, the system has a low/intermediate 
orbital eccentricity $e$ = 0.11 and P$_{\rm orb}$ = 250 days.
The \xr\ luminosity (0.1-10 keV) at its usual low state is about 
2$\times$10$^{34}$ ergs\,s$^{-1}$.
LTC12 found that during one of its \xr\ outbursts its \xr\ flux increased
and then subsided by a factor of five. The rise and decline each lasted
a year or so while its optical light curve showed little or no response.
As already noted, in contrast the largest \xr\ variations of \gc\ occur in
the form of its rapid ``flares." Its longer term variations do not exceed a 
factor of 2-3 at most, and these are correlated with optical flux.
X\,Per is the prototype of a small group of 
``persistent" Be slow X-ray pulsar systems.
These persistent systems are generally thought to be products of
``low kick" orbital perturbations resulting from a mild SN explosion.

  The \xr\ emissions from a Be-NS system are determined by details of a
wind flow from the Be star (or its disk) to a magnetized NS 
companion.\footnote{We use the term ``wind" throughout, in the understanding
that the outflow may not be due to the usual line-driven radiative wind 
occurring in most OB stars \citep[see e.g.,][]{Carciofi et al.2012}.} 
Magnetic field strengths of these pulsars are estimated to be of the
order 10$^{12}$\,G or even higher. Particles from the wind are ultimately
deposited onto the surface of a magnetosphere (defined below), where they
are guided into columns by the pulsar's field toward its magnetic poles.
We observe the \xr\ emission along these columns as {\it pulses} because the
poles are rotationally advected across our line of sight. The resulting
spectrum is typically optically thick. In the case of X\,Per the spectrum
can be fit well at high energies with a hard power law (index $n$ $\approx$2)
and at low energies by a relatively cool blackbody (e.g., LTC12). These
attributes are very different from those in the \gc\ spectrum.

\section{The propeller mechanism}
\label{prp}

In describing the geometry of particle accretion onto X-ray pulsars we
should first summarize the properties of three characteristic radii of
volumes around the NS.
The first is the well known Bondi-Hoyle radius, $R_B$,
within which wind particles from the Be star or its disk spiral 
toward the NS due to gravitation. They will deposit their energy
on the star's surface as X-rays or in its magnetosphere if the NS is highly 
magnetic. 

The second important size is of the magnetosphere, given by the Alfven radius 
$R_A$. This is the distance from the NS for which the decreasing magnetic 
pressure is balanced by the ambient ram pressure from the wind. 
Inside this radius the magnetic pressure wins and so the magnetosphere 
rotates rigidly with the NS and carries along embedded plasma. In the simple 
dipole case captured wind particles flow mainly toward the magnetic poles. 

The remaining important radius is the corotation radius, $R_C$, which is the
point at which the angular velocity of particles orbiting the star equals the 
NS rotational value. Inflowing magnetically channeled wind particles just 
beyond this radius are super-Keplerian and spiral outward to a larger orbit.

For typical Be-magnetic NS systems
the Alfven radius lies {\it interior} to the corotation radius. 
This permits wind particles to freely transit from their origin to the 
rotating magnetosphere and be captured by magnetic stresses at its surface. 
The kinetic energy of the particles is transformed chiefly to X-rays,
which typically display a power-law (nonthermal) spectrum. 
The particles are channeled along magnetic flux 
lines toward the NS.  The luminosity L$_x$ depends in part on the
accretion radius, which in this case is the much larger Bondi radius, $R_B$.

 For the case of \gc\ POT17 have considered an alternative propeller regime, 
for which $R_C$ $<$ $R_A$, which may occur for fast rotating pulsars.  
In this configuration the rigidly rotating magnetosphere extends
to a region where wind particle velocities are super-Keplerian, they can
fall no further than the magnetosphere's boundary. 
The centrifugal forces at the surface of the magnetosphere prevent particle
accretion such that most of them are deflected outwards. 
They subsequently accumulate within a shell bounded by $R_A$ $<$ $r$ $<$ $R_B$,
with most of them concentrated at the inner boundary.
The particles are then pushed by the upstream ram pressure of the wind
deep into the magnetosphere.
The surface shell remains optically thin and so it emits little UV or optical 
flux.
The accretion radius (and hence L$_x$) and temperature of the heated plasma 
is now partly determined by $R_A$, which in turn is much smaller than $R_B$. 
Notice that a substantial \xr\ luminosity and hard spectrum can result. 
A number of papers in the literature have suggested that the outburst or 
high state of short period BeXR systems occurs when the propeller mechanism 
temporarily ceases to operate and uninhibited accretion resumes 
\citep[see e.g.][and references therein]{Tsygankov et al.2016,Christodoulou2016,
Reig&Milonaki2016}. 
These outburst events occur on short timescales, at least in part as
response to surges in wind density, leading to brief
outbursts from direct accretion. Since $R_A$ shrinks in response to increases in
accretion density, the inequality is reversed, that is $R_A$ $<$ $R_C$, during
these outbursts. Again, this outburst behavior is in contrast to 
long term changes in the accretion phase forced by NS spin-down and binary 
evolution. These evolutionary considerations are discussed in the next section.

\section{Evolutionary timescales}

The POT17 proposal merges the above ideas into a putatively integrated
evolutionary scenario. These authors pictured \gc~ as a progenitor of a class 
of low to moderate eccentricity BeXR systems such as X\,Per. 
In their description of the \gc\ system the putative fast rotating NS is 
in the propeller mode and gradually loses rotational energy owing to a braking torque of the 
surrounding magnetosphere. This torque acts from radiative losses
and mechanical drag on the rotating magnetosphere by the wind. 
As the rotational angular velocity of the NS slows, the corotation radius $R_C$ 
increases, moves outside $R_A$, and direct accretion via gravitation can start. 
POT17 estimated that the propeller phase could last ``several 10$^5$ years 
or even longer."
Next POT17 advanced several predictions to adduce the applicability 
of their picture. 

We will comment on the POT17 predictions
and as well as on the applicability of their scheme in $\S$\ref{rebt}.
We review first the durations of the so-called ejector and propeller 
phases of the magnetic neutron star. These durations determine the sizes of the population of 
\gc-type systems, which one can compare with the number of known systems in 
our Galactic neighborhoods.  
In $\S$\ref{sppe} we will present a more complete evolutionary scheme than did 
POT17 for the transfer of wind particles to the putative NS. 

Our scheme has the sequence: \\
$ ejector \rightarrow propeller \rightarrow direct\, accretion \rightarrow 
spin\, equilibrium. $ \\

\noindent The ejector phase will be discussed in $\S$\ref{ejph}.
As implied, ``direct accretion" here simply refers to the flow of
wind particles to the surface of the NS in the case of no (or weak) 
magnetic field; as noted by POT17 the resulting L$_x$ can be very 
high because the full gravitational energy of the former wind particles 
is liberated near the small NS radius.  Spin equilibrium, 
which perhaps should be more accurately called a ``quasi-equilibrium,"
occurs when the Keplerian orbital velocity of the outer edge of the 
magnetospheric disk equals the corotation velocity at that point, 
as defined by \citet{Waters1989}.  The X-ray source then resides 
on the Corbet relation for BeXR systems \citep{Corbet1984}. Also note that 
if the accretion rate changes rapidly the NS system can transition back 
and forth between the propeller and ejector modes on short timescales.

\subsection{The velocity of the accreted material}
\label{digrs}

The mass accretion rate required to feed the NS in the propeller 
mode contrains the possible range of velocities of the accreted material at 
the Bondi radius. Importantly, as noted by POT17 the duration of the 
propeller phase critically depends on this velocity.

According to Eq. 15 of POT17, the mass accretion rate required to sustain 
an X-ray luminosity of $L_{32} \times 10^{32}\mathrm{erg\, s}^{-1}$ on a 
neutron star in propeller mode is  
\begin{equation}
\dot{M_{\rm x}}  \approx 8.3\times 10^{-11}\mu_{30}^{2/15}L_{32}^{4/5}\ M_\odot yr^{-1},\\
\label{mdot}
\end{equation}
where $\mu_{30}$ (which is $\mu/(10^{30}\mathrm{G\,cm}^3$), the typical 
dipole magnetic moment of the neutron star, is approximately unity, and 
assuming that the mass of the NS, $M_{\rm x}$ is 1.4\,$M_{\odot}$. 
The latter is larger than the 1.0\,$M_{\odot}$ value taken by POT17 
and is a more realistic value, as noted below; see also 
\citet[][]{Ozel et al.2012, Ozel&Freire2016}.
Since accretion occurs though 
wind capture $\dot{M_{\mathrm x}}$ is the Bondi accretion rate 
and $\dot{M_{\rm x}} = \piup \rho(r) V_{\mathrm{o}} \ R_{\mathrm B}^2$, with
the Bondi radius expressed as $R_{\mathrm B} = 2GM_{\rm X}/ V_{\mathrm o}^2$. 
Here we have retained the POT17 notation for parameters.
This includes $V_{\mathrm o}$ for the vectorial sum of the wind and orbital 
velocities to which should be quadratically added the sound speed in the 
accreted material $\approx 10 \times (T/10^{4}K)^{1/2}$ km\,s$^{-1}$. 
Our notation for parameters of the Be star will be an asterisk.

Assuming that the neutron star accretes matter from the 
intermediate latitude wind that is typical of an early B star, we can relate the mass 
accretion rate onto the NS to the total mass loss of the Be star. 
With $M_{*} = 15 M_{\odot}; M_{\mathrm x} = 1.4 M_\odot$ and 
$P_{\mathrm{orb}} = 203.59$d, $M_{*}$, the total mass loss rate of the 
primary wind, should be 
\begin{equation}
 \dot{M_{*}} = 1.2 \times 10^{-4} 
 (\frac{V_{\mathrm{o}}}{1000 \mathrm{km\,s^{-1}}})^{4}\ M_\odot yr^{-1},
\end{equation}
which is a few orders of magnitude above the total wind mass loss rate 
expected for a B0\,IV star, $\dot{M_{*}} \sim  10^{-8} \ M_\odot yr^{-1}$. 

From the foregoing, one can see that in order to capture enough material 
the orbit of the 
neutron star has to be nearly coplanar with the plane of the decretion disk 
and accrete low relative velocity and high density matter. The fraction of 
observed Be-shell stars led Porter (1996) to conclude that the disk opening 
angle is typically $\approx$ 5 degrees, although disk flaring at large radii 
may weaken this constraint. 

 The range of observationally permitted orbital inclinations depends on the 
assumed masses of the B0 star and NS.  Assuming the values given above
for the primary and secondary stars, combined with the observed 
radial velocity semiamplitude of 3.8 km\,s$^{-1}$ (SLM12a),  
implies a low orbital inclination on the order of $i\ \approx\ 
$29$^{\mathrm o}$, which is at variance with that of the decretion 
disk ($i = 42^{\mathrm o}$) derived from optical interferometry (SLM12a). 
Such a significant misalignment of the NS orbit with the disk plane 
can be expected to drive an X-ray flux modulation at the orbital period.

POT17 argue that the helium star electron capture channel 
could yield neutron stars with masses as low as 1$M_{\odot}$. 
In this case, a primary with a somewhat high mass of 16$M_\odot$ would indeed
allow a 1$M_{\odot}$ mass NS 
(their assumed mass values) 
to orbit close to the plane of the decretion disk. 
However, there is no observational evidence of such low mass ($\le$ 
1$M_{\odot}$) neutron stars. 
Although rather few reliable NS mass estimates exist for 
Be/X-ray systems, not to mention \gc -like systems, masses of the non-recycled 
NS+WD or NS+NS binaries - descendent of these systems - do not show evidence 
of a significant population of low mass neutron stars 
\citep[see e.g.,][]{Ozel&Freire2016}. 
In addition, the amount of accreted matter during the putative X-ray active 
stage of a few 10$^{6}$ yr is only $10^{-3}M_{\odot}$ at most. 
Therefore we will take the mass of the NS as constant over the entire period.

Optical spectroscopic observations of Be stars have established that the 
decretion disk is essentially Keplerian, with no detectable outflow
velocity \citep[see e.g.][and references therein]{Stee et al.2012}.
As already noted, in low eccentricity Be/X-ray binary systems such 
as \gc\ the disk is very efficiently truncated at the 3:1 
resonance radius, leaving a wide gap size between the edge of the disk 
and the neutron star (Okazaki \& Negueruela 2001).  
Additional numerical simulations by \citet{Okazaki&Negueruela2001} 
suggest that the infalling disk matter acquires velocities relative to the 
neutron star comparable to the sound speed and are at most of the order of a 
few tens of km\,s$^{-1}$. 
For these velocities the Bondi accretion radius is comparable to the radius of the Roche lobe. At the distance of the Roche radius of the accreting object in \gc, the free-fall velocity is $\approx$80\,km$\,s^{-1}$. 
Whatever the range of primary and secondary masses considered, the orbital velocity of the accreting object remains at nearly similar values of the order of 80-90 km\,s$^{-1}$ for low eccentricities.
It is therefore unlikely that the V$_{0}$ velocity in Eq. 14 of POT17  reaches 
values much above 100 km\,s$^{-1}$. Accordingly, durations of the propeller 
phase in excess of a few 10$^{5}$yrs are very unlikely.

\subsection{Spin period evolution}
\label{sppe}

\subsubsection{Ejector phase}
\label{ejph}

The very first phase of the neutron star history is the so-called ejector 
phase during which the ram pressure of the wind entering the gravitational 
influence of the neutron star at the Bondi radius is lower than the outgoing 
flux of electromagnetic waves and relativistic particles emitted by the 
magnetic neutron star (pulsar) \citep[see e.g., ][``PT12'']{Popov&Turolla2012}. 
This condition is expressed as $P_{\mathrm dyn} \leq \ P_{\mathrm PSR}$, 
with $P_{\mathrm dyn} = \rho(r) 
 V_{\mathrm{o}}^{2}$ and $P_{\mathrm PSR} = \dot{E} / (4 \piup R c) $. 
The power radiated by the slowing down pulsar due to dipole radiation is 
\begin{equation}
\dot{E} = 8 \piup^{4} B^{2} R_{\mathrm ns}^{6} \sin^{2}\alpha / (3 c^{2} P^{4}),
\end{equation}    
where $B$ is the magnetic polar dipole field, 
$R_{\mathrm X}$, the radius of the neutron star and $\alpha$ the 
inclination of the magnetic dipole from the rotation axis.

When does this ejector phase end? This question can be addressed by first
assuming the POT17-adopted values of density and velocity for the 
incoming matter at the Bondi radius sufficient to explain the observed 
X-ray luminosity in the propeller mode (eq. \ref{mdot}), and also by using
PT12's Eq. (1). We can then determine the spin period $P_{\mathrm{ee}}$ at 
which the ejector mode ends. This is

\begin{equation}
P_{\mathrm{ee}} \approx 0.48 \times 
 (\frac{V_{\mathrm{o}}}{100 
\mathrm{km/s}})^{-1/4} (\frac{B}{10^{12}})^{1/2} L_{32}^{-1/5} s,
\end{equation}
assuming $M_{\mathrm x} = 1.4 M_\odot$ and $\alpha$ = 90 degrees.\\

\noindent Next, following PT12, and assuming a canonical moment of inertia 
$I = 2/5 MR^{2}$ equal to 10$^{45}$ gm cm$^{2}$, the duration of the 
ejector phase may be computed as 
\begin{equation}
\tau_{\mathrm{ej}} \approx 3.6 \times (\frac{V_{\mathrm{o}}}{100 
\mathrm{km/s}})^{-1/2} (\frac{B}{10^{12}})^{-1} L_{32}^{-2/5}~~ Myr.
\end{equation}

\noindent Assuming factors of order unity for parameters in eq. 5,
$\tau_{\rm ej}$ should be a few million years. In addition, the putative
neutron star should have spun down already significantly during the 
ejector phase.

\subsubsection{Propeller phase}
\label{prpph}

As implied in $\S$\ref{prp}, the end 
of the propeller accretion mode occurs when the corotation radius, 
ever increasing because of NS spin-down,
reaches value close to the magnetosphere radius. 
Using Eq. 7  of POT17 for the magnetosphere radius and assuming a 
1.4\,$M_{\odot}$ neutron star implies that the high X-ray luminosity 
accretion mode switches on as soon as the spin period becomes longer than 
\begin{equation}
P_{\mathrm ep} = 8.6 \times \mu_{30}^{4/5}L_{32}^{-1/5} \ s.
\end{equation}

The braking torque acting on the neutron star in the propeller mode is given by Eq. 12 in POT17. Importantly, the torque is constant and does not depend on the difference between the critical and actual angular frequencies. Accordingly, the duration of the propeller phase is of the order of 
$\tau_{\mathrm {prop}} = 2 \piup (1/P_{\mathrm{ep}} - 1/P_{\mathrm {ee}})/\dot{\omega}$, with 

\begin{equation}
\dot{\omega} = -49\, I^{-1}\omega_{\rm B}^2 r_{\rm B}^3 C 
\frac{R_{\rm A}L_{\rm X}}{GM_{\rm X} V_{\mathrm{o}}},
\end{equation}

\noindent with $\omega_{\rm B}$ the orbital angular frequency and $R_{\rm A}$ 
the magnetosphere radius. 
The duration of the propeller phase is given by POT17 in their Eq. 14. 
It should be reiterated that some significant spin-down will have 
occurred during the relatively long ejector phase. 
Here we use $P_{*} = 0.48s$ as derived in our eq. 4, a value much longer that 
the probable birth spin period. Importantly,
this significantly further shortens the duration of the propeller phase. 
Using the Alfven radius in their Eq. 7, an average X-ray luminosity of 
5$\times$10$^{32}$ erg/s (SLM16) and $P_{\mathrm orb}$ = 203.5\,d, 
we can express the  actual duration of the propeller phase as 
\begin{equation}
\tau_{\mathrm {prop}} \approx 5 \times 10^{5} 
(\frac{V_{\mathrm{o}}}{100 \mathrm{km/s}})^{7} yr.
\end{equation} 

\noindent Note that this timescale is consistent with our estimate 
in $\S$\ref{digrs}, based on the range of velocities discussed there,
and that it is several times 
shorter than the duration of the ejector phase given in eq. 5.

\section{Our verdict: no propeller for \gC}
\label{rebt}

\subsection{Propeller systems in context of Be/X-ray binary evolution}
\label{relatn}

Several independent evolutionary arguments just discussed 
suggest that the propeller mechanism as proposed 
by POT17 cannot account for the space density of the \gc\ X-ray phenomenon.

First, as noted in $\S$\ref{digrs}
the velocity of the accreted material at the
Bondi radius is unlikely to be much greater than $\approx$ 100\,kms$^{-1}$. 
As we have seen, the velocity of the dense flow extracted from the outer 
edge of the decretion disk 
probably does not reach values larger than a few tens of km\,s$^{-1}$. 
In addition, we have shown that a large part of the neutron star spin-down 
may have already occurred during the ejector phase. From these considerations 
one can see that both parameters $P_*$ and $V_o$ in Eq. 14 of POT17 probably 
have values consistent with a propeller phase duration of only on the order 
of a few 10$^{5}$ yr.
It seems clear that the steep dependency of $T_{\mathrm {prop}}$ on 
relative velocity disallows the conclusion that the duration of the propeller 
phase can be as high as 1\,Myr. 
Second, although there may be a few systems in the propeller phase,
we consider it unlikely that a large population of NS/\gc -like propeller 
systems exists in the Galaxy. This is because of the rather short
duration of the propeller phase. We develop this point in the following.

Over 100 HMXBs are known in the Galaxy, among which $\approx$80\% are confirmed or 
candidate Be/X-ray systems \citep[][]{Liu et al.2006, Reig2011}. 
HMXB population models, e.g., by 
\citet[][``PV96'']{Portegies Zwart&Verbunt1996},
\citet[][``SL14'']{Shao&Li2014} predict the existence of $\approx$ 500 
Be/X-ray binaries in the Galaxy - a figure roughly consistent with the number 
of systems known, taking into account observational biases. With a formation 
rate of $\approx 5 \times 10^{5}$yr$^{-1}$, including effects of NS
birth kicks, the lifetime of X-ray Be systems is of the order of 
$\approx$10\,Myr (PV96).

Given the observed long Be/X-ray phase, comparable to the evolutionary time 
scale of an early type mass donor star, the number of progenitors of putative 
\gc\ propellers cannot be much higher than that of directly accreting systems.
Consider therefore the following contradiction.
Apart from a handful of nearby systems detected in low sensitivity all-sky 
surveys (HEAO-A1 and Rosat all-sky), the great majority of \gc-like systems
were discovered in XMM-Newton galactic surveys (SLM12a). 
Only a few percent of the Galaxy have been observed by XMM-Newton, 
and fewer still have been followed up in optical spectroscopy. 
Nebot et al. (2013) report the discovery of 4 new \gc-like objects at 
distances of $\approx$2\,kpc in a 4 deg $^{2}$ survey. 
Taken at face value, and even considering larger errors, 
this implies the presence of several 
thousand \gc-like objects in the Galaxy, a figure much larger than the
500 members predicted by PV96 and SL14 for the entire set of Galactic Be/X-ray 
binaries.

Conversely, one may compare the frequency of systems in the ejector phase to 
systems in the propeller phase. Collision of the relativistic pulsar 
wind with the stellar wind generates copious high energy emission across the 
X-ray to the $\gamma$-ray regime \citep[see e.g.,][]{Bogovalov 
et al.2008}. The spectral energy distribution of the binary 
Be/radio pulsar binary PSR B1259-63 (so far the only one known)
peaks in the 10-100 MeV energy range and 
extends up to more than 100\,GeV \citep[][]{Abdo et al.2011}. 
\citet[][``D13'']{Dubus2013} convincingly argue that most $\gamma$-ray binaries
are made of young radio pulsars embedded in the circumstellar material of a 
massive star. Only pulsars with spin-down luminosities above 
10$^{35}$ erg\,s$^{-1}$ are energetic enough to produce $\gamma$-rays, 
implying $\gamma$-ray life times of the order of 
6$\times 10^{5}$ yr (D13).\footnote{However, even for these systems
the NS star remains $\gamma$-ray quiet during the ejector phase.}
Even given the attendant observational biases, the low observed 
frequency of Galactic $\gamma$-ray binaries
(only this one Be/NS binary is known in the $\gamma$-ray active ejector state)
is consistent with the population of their 
wind-accreting BeXR binary descendants.  This provides further support for 
the conclusion that the propeller mechanism is unable to explain the number 
of \gc\ stars observed.

POT17 further assume that all \gc -like systems must have a low eccentricity, 
due to the low kick velocity imparted by the particular supernova mechanism 
assumade. However, there is no evidence for such a large population of systems 
existing before the propeller phase (ejectors), nor afterwards 
(classical Be/X-ray binaries).  In particular, all $\gamma$-ray ejector binaries have very eccentric orbits ($e \geq 0.35$, including PSR B1259-69). In addition, all young radio pulsars in binaries with massive companions that have terminated their $\gamma$-ray active phase but are still in the ejector phase also have very high eccentricities ($e \geq 0.58$; \citet[][``M05'']{Manchester2005}, D13) and spin-down times ($\tau_{sd} \approx 3-5 Myr$; M05), very consistent with the duration of the ejector phase computed in eq. 5. The range of eccentricities observed in post-SN systems is consistent with that observed in Be/X-ray binaries in which the vast majority of the systems have eccentricities $\geq 0.3$ (Reig 2011).
In addition, the great majority of Be-slow pulsar systems 
like X\,Per have noncircular orbits. 
Indeed, in the sample of eight long-period pulsar systems identified by
\citet[][]{Knigge et al.2012} none is in an almost circular ($e$ $\le$0.03) 
orbit.  If such systems are in noncircular orbits, while \gc\ is not, then these
two types of systems would appear to be members of two distinct populations. 

Apart from the \gc\ stars, there is no reason why a long-lived 
propeller stage {\it could not} exist in Be-NS binaries with larger 
eccentricities or inclined orbits as progenitors of the bulk of Be/X-ray 
systems.

\subsection{Our primary nonevolutionary objections to the Postnov et al. scenario}

POT17 do not address the critical details of the optical-UV-X-ray
variations or of the \xr\ flares. In fact, references in their
paper to ``ultraviolet" or ``spectroscopic variations" were not 
made and apparently not considered. As outlined above, the correlation 
and anti-correlation of various UV spectral lines and UV continuum with X-ray
variations means that the \xr\ emitting plasma is strongly influenced by the 
only nearby major source of UV and optical wavelength flux, the Be star.  
POT17's overlooking of this fact is our first major objection, apart from 
evolutionary issues just discussed.


Second, the POT17 description of what we call flares (or ``shots") 
was characterized as merely the short timescale end of a continuous
distribution of variabilities. In fact, the short flare decay 
rates are critical to a more focused consideraton as they necessarily 
imply particle densities of up to 10$^{15}$\,cm$^{-3}$ for the pre-flare 
parcels (SRC98).
Such high densities are hardly characteristic of POT17's hot shell 
but they {\it are} characteristic of the Be star's lower atmosphere, 
and realistically only of this site.  Insofar as these observations are not 
addressed, the POT17 picture is lacking.

  Third, the correlation of optical and \xr\ $\sim$ 70 day cycles mentioned
above for \gc\ has important implications. The existence of the red-tinged 
optical variation implies oscillations with a source that is cooler than 
the surface of the Be star yet still competing in its optical radiation 
with the Be star. The inner part of the decretion disk alone qualifies as
the secondary source of red-optical light. This means the dense inner region 
of the disk must somehow be associated with the creation of most or all 
of the hard \xr\ flux.

  Fourth, in 2010 the soft-\xr\ flux of \gc\ decreased relative to its hard 
flux. This occurred during a 44 day monitoring period that coincided with an 
optical outburst. This event added matter
to the line of sight column density. In other words, the hard X-ray source 
(which produces most of the soft X-ray flux as well) must lie close to
the outbursting Be star.  We stress here that the hot sources cannot
be placed anywhere else but on the surface of the Be star - for example, 
in the inner disk as the density is too low to be consistent with the 
short flare decays mentioned above.

Fifth, POT17 claimed the prediction of a 40-day time lag of X-ray 
behind optical signal and cited a MLS15 result as confirmation. This is 
a  misreading. Firstly, MLS15 found no lag at all. The ``one month" figure 
MLS15 quote refers to a generous upper limit, not an equality. 
Secondly, the optical/\xr\
time for any lag in Be-NS X-ray systems is not only the free-fall time into
the NS potential well but must also include the transit times from the stellar 
ejection and transits through the Be disk and to the secondary companion. 
One must look to actual empirical examples of a transit 
time. A literature search for such correlated outbursts by MLS15 reveals two 
cases of optical-\xr\ lag (one of them is X\,Per itself).  The 
optical/\xr\ lags in these cases, that is from  the Be outburst to a response
at the NS, have been observed to be about 4 years in both cases, 
\citep[e.g.,][]{Haubois et al.2012, Carciofi et al.2012}.
Thus, this is the lag timescale one can expect under the POT17 or other 
accretion scenarios.  
It is in disagreement with the observational result of MLS15, which again pointsto the X-ray emission of \gc\ arising from the vicinity of the Be star. 

Any one of these arguments is sufficient to vitiate, or at the very least 
seriously question, whether the propeller mechanism is applicable to
the \gc\ stars.

\subsection{Other criticisms of the Postnov et al. scenario}

To complement the foregoing, we rebut purported predictions made by POT17:

\begin{itemize}

\item The propeller/hot shell picture predicts a continuous range of 
temperatures for the heated plasma from an initial high value, $kT_{\rm hot}$ 
(POT17, eq. 3). Although we do not understand ourselves the origin 
of the secondary plasma components surrounding \gc\,~ the evidence for their 
existence as discrete structures (e.g., SLM12a, SLM12b) contradicts 
the finding by POT17 that the temperature distribution should be continuous.

\item The value of kT$_{\rm hot}$ derived by POT17 for the magnetosphere 
base is quoted as $\approx$27-32\,keV. Since this figure is actually based 
on a selection of particular NS mass and radius values by the authors among 
a range of possible values, it is not necessarily a prediction. 
The stated temperature also depends upon the unknown mass flow
rate impacting the Bondi sphere. 
Moreover, although a field strength of 10$^{12}$\,G is generally quoted,
the dispersion of these strengths among  X\,Per-like pulsars is not well
known. Given these uncertainties we doubt that the temperature according 
to their propeller model can be reliably predicted.

\item Similarly, we note that two propeller systems, 
4U\,01165+63 and V\,0332+53, that have been caught transitioning from a 
nonpropeller to propeller phase, exhibit soft, not hard, X-ray spectra 
during the latter phase \citep[][]{Tsygankov et al.2016}.
This is obviously contrary to expectation from the POT17 model.

\item  POT17 discuss a ``continuous" distribution of variability timescales 
from several seconds to a few days in $\gamma$\,Cas's \xr\ light curve. 
However, this is somewhat misleading. 
Although periodograms of \xr\ variations for \gc\ indeed vary
monotonically over these timescales,
a detailed inspection of several data series 
reveals departures from a simple or smooth frequency pattern, and
the degree of these departures varies from epoch to epoch (e.g., RS00).
Importantly, POT17's discussion implies 
that the rapid variations are part of a chaotic pattern of variations. 
This characterization is in stark contrast to their character as isolated, 
plainly visible sharp features in all high resolution time series. 
As noted already, for timescales $>$1 day there is also a significant peak 
due to the $\sim$70 day cycles in the low frequency region of the periodogram, 
Additional signal extends up to about one year. 

\item POT17 estimate an Emission Measure (EM) of $\approx$3.7$\times$10$^{54}$ 
cm$^{-3}$ for \gc\ from their model. 
Without indicating whether more optimistic selections of parameters 
are still reasonable, they added that the EM can be some ten times higher.
Available spectroscopic analyses of $\gamma$\,Cas indicate EM 
values of up to 3$\times$10$^{55}$\,cm$^{-3}$ (S04, SLM12a), or ten times 
above the best POT17 model. Similar values are obtained for HD\,110432 
\citep[][``LSM10'']{Torrejon et al.2012, Lopes de Oliveira et al.2010}.

\item  POT17 estimated the convective velocity of 1000 km\,s$^{-1}$ 
in their hot shell and they averred this to be an important prediction of
\xr\ line broadening. 
Actually, typical turbulence values measured in lines in high resolution 
\xr\ spectra of \gc\ are 300-500 km\,s$^{-1}$ (S04 and SLM12a).  
POT17 overlooked these findings.  The signal to noise ratio in Fe lines 
is generally too poor to make a reliable determination (LSM10),
and interpretational issues occur for these lines as well. 
Also, SRC98 showed that velocities of $\gtrsim$1300 km\,s$^{-1}$ 
(the expulsion as well as thermal velocity) could be predicted in their 
description of exploding surface flare parcels.

\item POT17 state that the observation of the Fe fluorescent feature 
is another prediction of their picture. However, such features are 
present in many types of \xr\ binary systems, regardless of the status of
a degenerate companion and \xr\ generation process.

\item An important characteristic of the POT17 model is that 
the orbital and Be disk planes are aligned.
As noted in $\S$\ref{digrs}, a significant tilt of the orbital plane would 
cause an X-ray flux modulation. Such a modulation is not seen (MLS15).~
Even a small kick may well move the NS out from its former orbital plane. 
Notice that there is no reason why a kick experienced by a NS explosion
should be directed within the orbital plane.  Since the POT17 hypothesis 
requires that {\it all} \gc\ systems have propellers, it follows that 
the kicks of all of them be confined to their NS's equatorial planes. 
This would be a highly unlikely series of events.  

\item A salient feature of \gc-like stars is their narrow range of X-ray 
luminosities. This would imply in the Postnov et al. scenario that in all 
\gc\ systems, the NS should orbit at nearly the same distance from the Be 
star.  Their decretion disks should also have similar extents and densities.
We have already discussed already the
improbable small ranges in orbital eccentricity and inclination implied.

\item In the HR Diagram the \gc\ stars are confined approximately to 
spectral types B0-B1.5 and luminosity classes III-V.  One can expect their 
progenitors on the ZAMS to be late-type O stars. However, the domain of
the persistent BeXR NS systems is somewhat larger. The persistent systems
could not be expected to evolve to as narrow a spectral type domain as
the \gc\ stars occupy.

\end{itemize}

\section{Final considerations}

In contrast to the propeller mechanism proposed by POT17, RS00, SLM16,
and other studies cited therein have
led to the interpretation that the generation of hard X-rays from \gc\ are
caused by an interaction of magnetic fields from the Be star and its decretion 
disk. Our initial picture has evolved with the accumulation of new datasets 
and is particularly informed by analyses of data from a number of 
multiwavelength observational campaigns. As noted, the light curves and
spectra we analyzed consist of simultaneous X-ray/UV from the 1996 campaign, 
as well as some 19 seasons of robotic two-color APT photometry, most of 
which were contemporaneous with X-ray monitorings. The latter  include the 
long term {\it RXTE} All Sky Monitor program.
An assessment of these combined datasets demonstrates that it is futile 
to construct {\it any} paradigm for the origin of the hard \xr\ that 
does not include detailed analysis of concomitant optical 
and UV variations. \\

  As noted by POT17, no magnetic signatures have been reported by
spectropolarimetry in $\gamma$\,Cas.  It is very
difficult to detect spectropolarization signatures  in the
very broadened lines of this star. Moreover, polarimetric techniques are 
tailored to the detection of magnetic dipoles, and other aspects of the star's
 behavior strongly suggest that any surface field cannot have a simple and
hence easily detectable topology.
Indeed, if the topology were simple the star's UV lines would be expected
to show variations characteristic of a magnetic Bp star, and they do not.
In recent years the periodic magnetic signature in the star's optical 
light curve has disappeared (HS12), rendering a polarimetric nondetection 
moot at the present time.

POT17 characterize the magnetic interaction scenario for \gc\ as 
``entirely phenomenological at present and lacking in predictive power." 
Actually, this is not quite true. For instance, in our interaction picture
hard \xr\ production would cause clear changes in the attributes of the hard
X-ray production, such as a disappearance of flaring, the value of k$T_{hot}$, 
abundances determined from X-ray lines, and the Fe\,K fluorescence feature.
The optical/UV migrating subfeatures and UV ``dips" might disappear as well.
The larger point is that the interaction picture connects the inner disk
conditions explicity with the X-ray production.  Since the disk disappeared
already in the early 20th century, we can look forward to it to disappear 
again and such tests to proceed.
Otherwise, it is certainly true that continued observations have
continually brought new surprises and necessitated changes to any
posited model for the hard \xr\ generation.

In all, we would say the ``phenomenological" aspect of our 
magnetic interaction scenario is an expression of its adherence to 
a broad array of multiwavelength observations as well as being
internally consistent.  In fact, it seems as 
fair to characterize the POT17 evolutionary scenario as being dependent 
on critical assumptions about the mass loss rate of the Be/Be-disk system
(by hypothesis the ``disk wind") and the amount of mass available for 
accretion onto a degenerate companion. POT17 and we agree that this rate 
is not well known - indeed scant progress has been made on this hard 
to determine parameter.

We believe the POT17 propeller scheme fails on the basis of the many 
considerations we have discussed.

\section*{Acknowledgements}

The authors wish to thank Dr. Chris Shrader for several edifying discussions.
We also heartily thank comments by the referee, Dr. Dimitris Chrisodoulou,
for clarifying important details pertaining to the propeller mechanism.
R.L.O. was supported by the Brazilian agencies CNPq 
(Universal Grants 459553/2014-3 and PQ 302037/2015-2) and INCTA (CNPq/FAPESP).


\bsp	
\label{lastpage}

\begin{thebibliography}{99}

\bibitem[\protect\citeauthoryear{Abdo etal.}{2011}]{Abdo et al.2011}
Abdo, A. A., Ackermann,M, Ajello, M., et al. 2011, ApJ, 736l, 11A

\bibitem[\protect\citeauthoryear{Balbus \& Hawley}{1991}]{Balbus&Hawley1991}
Balbus, S. \& Hawley, J. F. 1991, ApJ, 376, 214B

\bibitem[\protect\citeauthoryear{Bogovalov et al.}{2008}]{Bogovalov et al.2008}
Bogovalov, S. V., Khangulyan, D. V., Koldoba, A. V., et al., MNRAS, 387, 63B 

\bibitem[\protect\citeauthoryear{Carciofi et al.}{2011}]{Carciofi et al.2011} 
Carciofi, A. C., Bjorkman, J. E., Otero, S. A., et al. 2011, IAU Symp. Vol. 272,
ed. C. Neiner, 325

\bibitem[\protect\citeauthoryear{Carciofi et al.}{2012}]{Carciofi et al.2012}
Carciofi, A. C., Bjorkman, J. E. Otero, S., et al. 2012, ApJL, 744, L15 

\bibitem[\protect\citeauthoryear{Christodoulou et al.}{2016}]{Christodoulou2016} Christodoulou D.~M., Laycock S.~G.~T., Yang J., Fingerman S., 2016, ApJ, 829, 30 

\bibitem[\protect\citeauthoryear{Corbet}{1996}]{Corbet1996} Corbet R.~H.~D., 
1996, ApJ, 457L, 31C 

\bibitem[\protect\citeauthoryear{Corbet}{1984}]{Corbet1984} Corbet R.~H.~D., 1984, A\&A, 141, 91 

\bibitem[\protect\citeauthoryear{Dubus}{2013}]{Dubus2013}
Dubus, G. 2013, A\&ARv, 21, 64D (D13)

\bibitem[\protect\citeauthoryear{Eckstrom etal.}{2012}]{Eckstrom, et al.2012}
Eckstr\"om, S., Georgy, C,, Eggenberger, P., et al. 2012, A\&A, 537 A146E

\bibitem[\protect\citeauthoryear{Hamaguchi et al.}{2016}]{Hamaguchi et al.2016}
Hamaguchi, K., Oskinova, L., Russell, C., et al. 2016, ApJ, 832, 140H

\bibitem[\protect\citeauthoryear{Haubois et al.}{2012}]{Haubois et al.2012}
Haubois, X., Carciofi, A. C., Rivinius, T., et al. 2012, ApJ, 756, 156

\bibitem[\protect\citeauthoryear{Henry \& Smith}{2012}]{Henry&Smith2012}
Henry, G. W. \& Smith, M. A. 2012, ApJ, 760, 10H (HS12)

\bibitem[\protect\citeauthoryear{Knigge, C. et al.}{2012}]{Knigge et al.2012}
Knigge, C., Coe, M. J., \& Podsladlowski, P., 2012, Nature, 479, 372K

\bibitem[\protect\citeauthoryear{Liu et al.}{2006}]{Liu et al.2006}
Liu , Q. Z., van Paradijs, J., \& Van den Heuvel, E. P. J. 2006, A\&A, 455, 1165L

\bibitem[\protect\citeauthoryear{Lopes de Oliveira et al.}{2010}]
{Lopes de Oliveira et al.2010}
Lopes de Oliveira, R, Smith, M. A., \& Motch, C. 2010, A\&A, 512, A22L (LSM10)

\bibitem[\protect\citeauthoryear{Lutovinov et al.}{2012}]{Lutovinov et al.2012}
Lutovinov, A., Tsygankov, S., \& Chernyakova, M. 2012, MNRAS, 423, 1978L (LTC12)


\bibitem[\protect\citeauthoryear{Manchester}{2005}]{Manchester2005}
Manchester, R. N. 2005, Ap\&SS, 297, 101M (M05)

\bibitem[\protect\citeauthoryear{Motch et al.}{2015}]{Motch et al.2015}
Motch, C., Lopes de Oliveira, R., \& Smith, M. A. 2015, ApJ, 806, 177M (MLS15)

\bibitem[\protect\citeauthoryear{Okazaki \& Negueruela}{2001}]
{Okazaki&Negueruela2001}
Okazaki, A., \& Negueruela, I. 2001, A\&A, 377, 161O

\bibitem[\protect\citeauthoryear{Okazaki}{2001}]{Okazaki2001}
Okazaki, A. 2011, priv. commun.

\bibitem[\protect\citeauthoryear{Ozel et al.}{2012}]{Ozel et al.2012}
\"Ozel, F., Psaltis, D., Ramesh, N., et al. 2012, A\&A, 757, 55O

\bibitem[\protect\citeauthoryear{Ozel \& Freire}{2016}]{Ozel&Freire2016}
\"Ozel, F., \& Freire, P. 2016, ARA\&A, 54, 401O
\bibitem[\protect\citeauthoryear{Ozel \& Freire}{2016}]{Ozel&Freire2016}
 \"Ozel, F., \& Freire, P. 2016, ARA\&A, 54,401O ans

\bibitem[\protect\citeauthoryear{Popov \& Turolla}{2012}]{Popov&Turolla2012}
Popov, S.B., \& Turolla, R. 2012, MNRAS, 421, L127 (PT12)

\bibitem[\protect\citeauthoryear{Portegies Zwart \& Verbunt}{1996}]{Portegies Zwart&Verbunt1996}
Portegies Zwart, S., \& Verbunt, F. 1996, A\&A, 309, 179P (PV96)

\bibitem[\protect\citeauthoryear{Porter}{2017}]{Porter2017}
Porter, J. M. 1996, MNRAS, 280L, 31P

\bibitem[\protect\citeauthoryear{Postnov et al.}{2017}]{Postnov et al.2017}
Postnov, K., Oskinova, L., \& Torrej\'on, J. M. 2017, MNRAS, 465L, 119P (POT17)

\bibitem[\protect\citeauthoryear{Reig}{2011}]{Reig2011}
Reig, P. 2011, Ap\&SS, 332, 1R

\bibitem[\protect\citeauthoryear{Reig \& Milonaki}{2016}]{Reig&Milonaki2016} Reig P., Milonaki F., 2016, A\&A, 594, A45 

\bibitem[\protect\citeauthoryear{Robinson \& Smith}{2000}]{Robinson&Smith2000}
Robinson, R. D., \& Smith, M. A. 2000, ApJ, 540, 474R (RS00)

\bibitem[\protect\citeauthoryear{Robinson et al.}{2002}]{Robinson et al.2002}
Robinson, R. D., Smith, M. A., \& Henry, G. W. 2002, ApJ, 575, 435R (RSH02)

\bibitem[\protect\citeauthoryear{Shao \& Li}{2014}]{Shao&Li2014}
Shao, Y., \& Li, X.-D. 2014, ApJ, 796,37S (SL14)


\bibitem[\protect\citeauthoryear{Smith et al.}{2004}]{Smith et al.2004}
Smith, M. A., Cohen, D. H., Gu, M. G., et al. 2004, ApJ,600, 972S (S04)

\bibitem[\protect\citeauthoryear{Smith et al.}{2012a}]{Smith et al.2012a}
Smith, M. A., Lopes de Oliveira, R., \& Motch, C., et al. 2012, A\&A, 540, A53S (SLM12a)

\bibitem[\protect\citeauthoryear{Smith et al.}{2012b}]{Smith et al.2012b}
Smith, M. A., Lopes de Oliveira, R., \& Motch, C. 2012b, ApJ, 755, 64S (SLM12b)

\bibitem[\protect\citeauthoryear{Smith et al.}{2016}]{Smith et al.2016}
Smith, M. A., Lopes de Oliveira, R., \& Motch, C., 2016, AdSpR, 58, 782S
(SLM16) 

\bibitem[\protect\citeauthoryear{Smith \& Robinson}{2003}]{Smith&Robinson2003}
Smith, M. A., \& Robinson, R. D. 2003, in ``Interplay of Periodic, Cyclic,
and Stochastic Variability," ASP Conf. Ser. 292, 263S (SR03)

\bibitem[\protect\citeauthoryear{Smith et al.}{1998a}]{Smith et al.1998a}
Smith, M. A., Robinson, R. D., Corbet, R. H. 1998a, ApJ, 503, 877S (SRC98)

\bibitem[\protect\citeauthoryear{Smith et al.}{1998b}]{Smith et al.1998b}
Smith, M. A., Robinson, R. D., Hatzes, A. P.. 1998b, ApJ, 508, 945S 

\bibitem[\protect\citeauthoryear{Stee et al.}{2012}]{Stee et al.2012}
Stee, Ph., Delaa, O., Monnier, J., et al. 2012, A\&A, 545A, 59S

\bibitem[\protect\citeauthoryear{Torrej\'on et al.}{2012}]{Torrejon et al.2012}
 Torrej\'on, J. M., Schulz, N. S., \& Nowak, M. A. 2012, ApJ, 750, 75T

\bibitem[\protect\citeauthoryear{Tsygankov et al.}{2016}]{Tsygankov et al.2016}
Tsygankov, S., Lutovinov, A., Doroshenko, V., et al. 2016, A\&A, 593A, 16T

\bibitem[\protect\citeauthoryear{Waters \& van Kerkwijk}{1989}]{Waters1989} Waters L.~B.~F.~M., van Kerkwijk M.~H., 1989, A\&A, 223, 196 
\end{thebibliography}
\end{document}